\def\year{2022}\relax
\documentclass[letterpaper]{article} 
\usepackage{aaai22}  
\usepackage{times}  
\usepackage{helvet}  
\usepackage{courier}  
\usepackage[hyphens]{url}  
\usepackage{graphicx} 
\usepackage{natbib}  
\usepackage{caption} 
\DeclareCaptionStyle{ruled}{labelfont=normalfont,labelsep=colon,strut=off} 
\usepackage{algorithm}
\usepackage{algorithmic}
\usepackage{amsmath}
\usepackage{longtable}
\usepackage{comment}

\usepackage{newfloat}
\usepackage{listings}
\floatstyle{ruled}
\newfloat{listing}{tb}{lst}{}
\floatname{listing}{Listing}
\title{Blended Bots: Infiltration through Identity Deception on Social Media}
\author{
    Samantha C. Phillips\textsuperscript{1}, Lynnette Hui Xian Ng\textsuperscript{1}, Kathleen M. Carley\textsuperscript{1}
}
\affiliations{
    \textsuperscript{\rm 1} Software and Societal Systems, Carnegie Mellon University\\
    Pittsburgh, PA, United States
}

\begin{document}

\maketitle

\begin{abstract}
Bots are automated social media users that can be used to amplify (mis)information and sow harmful discourse.  
In order to effectively influence users, bots can be generated to reproduce human user behavior.  
Indeed, people tend to trust information coming from users with profiles that fit roles they expect to exist, such as users with gender role stereotypes.  
In this work, we examine differences in the types of identities in profiles of human and bot accounts with a focus on combinations of identities that represent gender role stereotypes.  
We find that some types of identities differentiate between human and bot profiles, confirming this approach can be a useful in distinguishing between human and bot accounts on social media. 
However, contrary to our expectations, we reveal that gender bias is expressed more in human accounts than bots overall.
Despite having less gender bias overall, we provide examples of identities with strong associations with gender identities in bot profiles, such as those related to technology, finance, sports, and horoscopes.
Finally, we discuss implications for designing constructive social media bot detection.
\end{abstract}

\section{Introduction}
Social media provides wider and faster access to people and information around the world. 
Concerningly, affordances of social media, like anonymity, are leveraged by nefarious actors to spread misinformation and harmful narratives among these online communication spaces \citep{kumar2018false,bradshaw2018challenging}. 
These aims can be operationalized using automated accounts, that is, bots \cite{ferrara2016rise}.
While bots are often useful (e.g., in crisis settings \cite{hofeditz2019meaningful}) or benign, they can also be used to deceive and manipulate users \cite{chang2021social}.
To effectively manipulate users, covert bots may attempt to blend in as human accounts through presentation and behavior \cite{wischnewski2022agree,freitas2015reverse}.
Bots that aim to imitate human behavior are often referred to as ``social bots'' \cite{boshmaf2011socialbot,stieglitz2017socialbots,stieglitz2017social,ferrara2016rise}.

One common method social bots use to reproduce human account behavior is to mimic user bios. 
A ``bio" on a social media platform is a brief description of the account or its owner \cite{madani2023measuring}. 
Users can select personal identifiers \cite[any phrase that describes the owner of the social media account: social identities, preferences, activities, and affiliations][]{pathak2021method} to put in their profile \cite[``self-presentation''][]{marwick2011tweet} that they believe are advantageous given their goals and perceived audience.
It follows that bots aiming to impersonate human accounts can select certain identities to include in their self-presentation, and therefore affiliate themselves with their target community \cite{ng2022cross}.

Studies show that people tend to respond positively to information they are familiar with due to cognitive biases like familiarity bias \cite{greul2023does}.
Familiarity is associated with increased trust in deceptive content \cite{zahari2019role} and fellow social media users \cite{shareef2020group}.
Therefore, we propose that social media users may be less likely to classify another user as a bot if their self-presentation contains expected combinations of identities and attributes, that is, common stereotypes.
Gender role stereotypes are common combinations of identities that are pervasive in news media \cite{armstrong2012global}, entertainment \cite{ward2020media} and AI-generated content \cite{kotek2023gender}.
Characterizing combinations of identities in the profiles of bot accounts informs training materials for both human and machine detection of bots \cite{kenny2023improving,ng2023botbuster}.




In this paper, we provide a large-scale quantitative comparison of identities in user profiles of accounts detected as bots and not detected as bots.
We focus this analysis on gender role stereotypes by examining co-occurring identities with gender identities in bios.
As such, we address the following research questions:
\begin{quote}
    \textbf{RQ1}: What types of identities are most and least prevalent in bios of bot and human users? \\
    \textbf{RQ2}: How do bot and human accounts present gender in their bio?\\
    \textbf{RQ3}: Is there more or less gender bias expressed in bot or human user account bios?  What types of identifiers are the most and least associated with gender identities in bot accounts?
    \\
    \textbf{RQ4}: What are the strongest associations with male and female gender identities in bot accounts?
\end{quote}
To address these research questions, we first employ a bottom-up approach to extracting identifiers from a sample of approximately 50M X (formerly Twitter) users.
We then use 23 categories of identifiers including family, religion, age, ethnicity, and technology \cite{phillips2024sorting}.
In RQ1, we confirm there are types of identities that are used disproportionately in profiles of human accounts (family, religion, school, occupation, activism) and types largely used by bots (relationship status, sex, technology, finance).

In addition, we categorize gender identities into eight subtypes: (female/male) pronouns, direct terms (e.g., woman, lady), and family roles, as well as transgender and non-binary/other.
In RQ2 we find that human account bios contain more female terms (e.g., woman, girl) and male pronouns; and used more male and female family roles. Bots, on the other hand, are more likely to include a male term in their bio (e.g., man, boy, gentleman), suggesting limited use of gender presentation to assimilate online.

Using the co-occurrences of extracted identifiers in bios, we generate networks for bot and human users.
We then measure the difference in weighted log-odds with informative Dirichlet priors between male and female identities.
That is, we calculate the difference in co-occurrence of each identity with male and female identities in bios while accounting for the overall use of the identity.
Contrary to our initial assumptions, in RQ3, we show humans display more bias along gender than bots overall and for the majority of identifier types.
Finally, in RQ4, we delve into specific examples of gender bias in bot accounts.

\subsection{Contributions}

Within this paper, we make the following contributions: 
\begin{enumerate}
    \item We describe a strategy for social bots to mimic human users by including common combinations identities in their profile, such as gender stereotypes.
    \item We empirically demonstrate differences in identity use and gender biases by human and bot accounts using a large dataset of 50.1M X user profiles.  
    \item Our analyses reveal identities in bot profiles that reflect gender biases, especially related to finance and technology.
    \item We discuss how the uncovered distinctions between bot and human accounts from this work may be useful in automated and manual bot detection training.
\end{enumerate}


\section{Related works}

\subsection{Deception by Social Bots}
Social bots have been observed using deception to effectively spread disinformation \cite{varol2019deception}, and promote propaganda \cite{jones2019gulf}.  
In response to concerns about the harmful effects of bots, many bot detection methods have been developed to differentiate real and fake accounts using user meta data and behavior \cite{ng2023botbuster,morstatter2016new,varol2017online,beskow2018bot}.
Bots also have been shown to use specific linguistic techniques in their messages \citep{addawood2019linguistic}. 

Relatedly, identity deception is prevalent on social media \cite{10.1145/3446372,van2018cyber}.
Identities in self-presentation can change how people perceive that account.
For example, there is evidence that bots presented as female are more likely to be judged as human \cite{borau2021most}.
In other contexts, bots that present as Black are judged as more trustworthy \cite{goble2016impact}.

Humans are generally inept at effectively distinguishing between bot and human accounts \cite{kenny2024duped}, generating feelings of uncertainty and lack of control. In these contexts, people are more likely to rely on their (biased) intuitions \cite{tversky1974judgment}.
Therefore, people rely on overt signals of human-ness like holding certain social identities or stereotypes. 
In other words, they use familiarity with the content to distinguish credible and incredible information and sources \cite{greul2023does,zahari2019role,shareef2020group}.
This work examines a pathway for bots to gain credibility and infiltrate online groups through use of commonly associated identifiers concurrently with gender signals, as well as selection of identifiers in their profile more broadly.

\subsection{Measuring Gender Biases on Social Media}
Research on gender bias in language and, more specifically, descriptions of bios have been prominent.  
Numerous works show pervasive gender biases in mainstream media \cite{asr2021gender,shor2019large}, Wikipedia \cite{graells2015first}, social media \cite{usher2018twitter,marjanovic2022quantifying}, and search engines \cite{wijnhoven2021search}.
Therefore, we expect gender bias in social media content.

Gender bias can be operationalized as the extent to which individual words and types of language are used by each gender, or to describe the gender \cite{graells2015first,guilbeault2024online}.
Hence, gender bias is determined by the association between words of identifiers and gender identities.

There are two broad approaches to measure associations between words.  
The traditional approach is to use semantic networks where nodes are words and edges denote co-occurrence within a window of text.
For example, previous research measured associations between words and gender terms using pointwise mutual information \cite{graells2015first}.
However, pointwise mutual information over-weights scarcely used words and quantifies association without considering the other side.  
Although not directly applied to gender associations, work from the late 2000s developed models to compare the associations of terms in two or more sets of documents that include some form of Bayesian prior \cite{monroe2008fightin}.
For example, informative Dirichlet priors allow researchers to account for overall use of a word in the corpus when measuring associativity.
In this way, the models proposed by Monroe and colleagues provide a computationally inexpensive, interpretable approach that avoids over-weighting of high- or low-frequency words.

Alternatively, in recent years, use of embeddings models to represent words in a high-dimensional space has become a popular way to assess associations between gender groups and words \cite{charlesworth2022historical,guilbeault2024online}.
These methods typically involve fine-tuning pretrained embeddings models using the texts of interest.  However, it can be difficult to assess the extent to which the embeddings are based on the fine-tuning or initial training dataset.

Furthermore, there is evidence that these models innately contain gender bias \cite{bhardwaj2021investigating,kotek2023gender}.
Previous work demonstrates that fine-tuned models largely preserve the structure of the original model \cite{zhou2021closer}.  
For these reasons, it may be challenging to determine what, if any, gender bias in the fine-tuned model can be attributed to.
Therefore, it is not ideal for comparison of the degree of gender bias of the same terms in different datasets. 
We could train a word embedding model from scratch, like Word2Vec, but we are still faced with possible weighting issues based on the association measure selected.
On the other hand, we know directly using co-occurrences of identities are necessarily sensitive to differences in language use and there are network-based association measures that account for overall use.



\section{Data}
This work uses the meta data of a sample of 50.1M X users collected via Twitter API in June 2017.
These users are a sample of the most active users in 2010-2015, identified using Twitter's Decahose stream.  
In pre-processing, we removed accounts with no bio or non-English bios.

\section{Methods}

\subsection{Identity pre-processing}

For each user in each time period, we extracted phrases from bios using the method developed by \citet{pathak2021method}. Their work develops the concept of a ``personal identifier''---roughly, a phrase that signals identity---and proposes a simple method to extract personal identifiers from text. However, because their method generates noisy phrases, we opt to develop methods to post-process the output of their model, as well as methods to subsample to only frequent phrases.

From the 11,141 phrases used by at least 1,000 users, we extract 3,670 identifiers using labels from \cite{phillips2024sorting}.
We evaluate identifiers categorized as: culture, nationality/location/ethnicity/race, occupation/industry, sports, school/subjects,  technology/gaming/cars, (social) media, horoscope, travel, religion, outdoors/nature/animals, business/finance, activism, disability/(mental) illness, sex, military, age, family, values, relationship status, sexuality/gender.  
A clarifying point: culture refers to music, art, food, fashion, and other cultural preferences and roles. 
The category labels are also from \cite{phillips2024sorting}.
Table \ref{tab:labels} provides the identifiers in each category.

\subsubsection{Gender identities}

Of the selected identifiers, we labeled those that explicitly reference gender; see Table \ref{table1}.  All authors agreed on gender labels.

\begin{table*}[t]
\centering
\begin{tabular}{lp{14cm}}
Category & Identities \\ 
\hline 
Female - pronouns &  she, her, hers\\
\hline 
Female - roles & woman, girl, lady, mrs, female,  camgirl, happy girl, fangirl, small town girl, baby girl, babygirl, businesswoman, country girl, blackgirlmagic, island girl, nasty woman, girly, crazy cat lady \\
\hline 
Female - family &  sister, mummy, nana, aunt, momma, mom, stepmom, full time mommy, supermom, mompreneur, grandma, housewife, mum, mother, devoted wife, teammommy, stay at home mom, house wife, aunty, sahm, super mom, grandmother, mommy, big sister, being a mom, sis, mama, wife, motherhood, new mom, wifey, happy wife, granny, daughter, auntie, grammy, full time mom  \\
\hline 
Male - pronouns &  he, him, his \\
\hline
Male - roles & gentleman, man, boy, mr, guy, male, country boy, renaissance man, outdoorsman, handyman, sportsman, cameraman, businessman, simple man, common man, salesman, fisherman, it guy, all round good guy, all around good guy, simple guy, all around nice guy, all round nice guy, nice guy \\
\hline
Male - family & papa, lucky husband, father, devoted husband, uncle, proudfather, brother, godfather, grandfather, happy husband, son, new dad, dad, grandad, hubby, nephew, full time dad, grandpa, daddy, husband \\
\hline
Transgender & transgender, trans, ftm, mtf \\
\hline
Non-binary/other & genderfluid, non-binary, nb \\
\hline
\end{tabular}
\caption{Gender identities terms.}
\label{table1}
\end{table*}

\subsection{Associativity of Identities with Gender}

We then measure the alignment of each identifier along gender dimension (female - male) by calculating the difference in weighted log-odds ratios with informative Dirichlet priors between the two groups \cite{monroe2008fightin} using the tidylo R package\footnote{\url{https://github.com/juliasilge/tidylo}}. 
That is, the usage difference of identifier $w$ between two groups $i$ and $j$ at time $k$ is calculated by
\begin{align}
    \delta_{kw}^{(i-j)} &= log\left[\frac{y_{kw}^{(i)} + \alpha_{kw}^{(i)}}{n_k^{(i)} + \alpha_{k0}^{(i)} - y_{kw}^{(i)} - \alpha_{kw}^{(i)}} \right] \\
    &- log\left[\frac{y_{kw}^{(j)} + \alpha_{kw}^{(j)}}{n_k^{(j)} + \alpha_{k0}^{(j)} - y_{kw}^{(j)} - \alpha_{kw}^{(j)}} \right]
\end{align}
where $y_{kw}^{(i)}$ is the number of users that used an identifier in group $i$ (e.g., left identifiers) and identifier $w$ in time period $k$, and $n_k^{(i)}$ is the number of users that used an identifier in group $i$.  The informative prior is $\alpha_{kw}^{(i)} = $ \textbf{y} $\cdot \frac{\alpha_0}{n}$, where $\alpha_{0} = \sum_w \alpha_w$, $\alpha_{k0}^{(i)} = \sum_w \alpha_w^{(i)}$, \textbf{y} $\sim$ Multinomial($n, \boldsymbol\pi$) are the counts in the entire corpus, $n = \sum_w y_w$ and $\boldsymbol\pi$ is a $W$-vector of multinomial probabilities.

We can then divide by the standard deviation of $\delta_{kw}^{(i-j)}$ to obtain z-scores of the log-odds-ratios:
\begin{equation}
    \zeta_{kw}^{(i-j)} = \hat{\delta}_{kw}^{(i-j)} / \sqrt{\sigma^2 (\hat{\delta}_{kw}^{(i-j)})}
\end{equation}

\subsection{Bot detection}
We used BotBuster, a mixture-of-experts supervised machine learning architecture to differentiate bots and humans \cite{ng2023botbuster}. This algorithm has been built and tested on a variety of datasets, ranging from elections-based data, fake follower bots and disinformation bots. In BotBuster, bot probabilities of different data pillars are aggregated together to classify the user. This dataset contains user names, screen names and description but does not contain the text of tweets. Therefore, only the BotBuster experts for user meta data are activated and aggregated together. BotBuster returns the probability of the account being a bot ($P(bot)$), and the corresponding probability of the account being a human ($P(human)$). The account is classified as a bot if $P(bot)>P(human)$. Within our dataset, about 33\% of users are classified as bots.

\section{Results}

\subsection{Comparison of types of identities in bios of bot and human users (RQ1)}

\begin{figure*}[t]
    \centering
    \includegraphics[width=\textwidth]{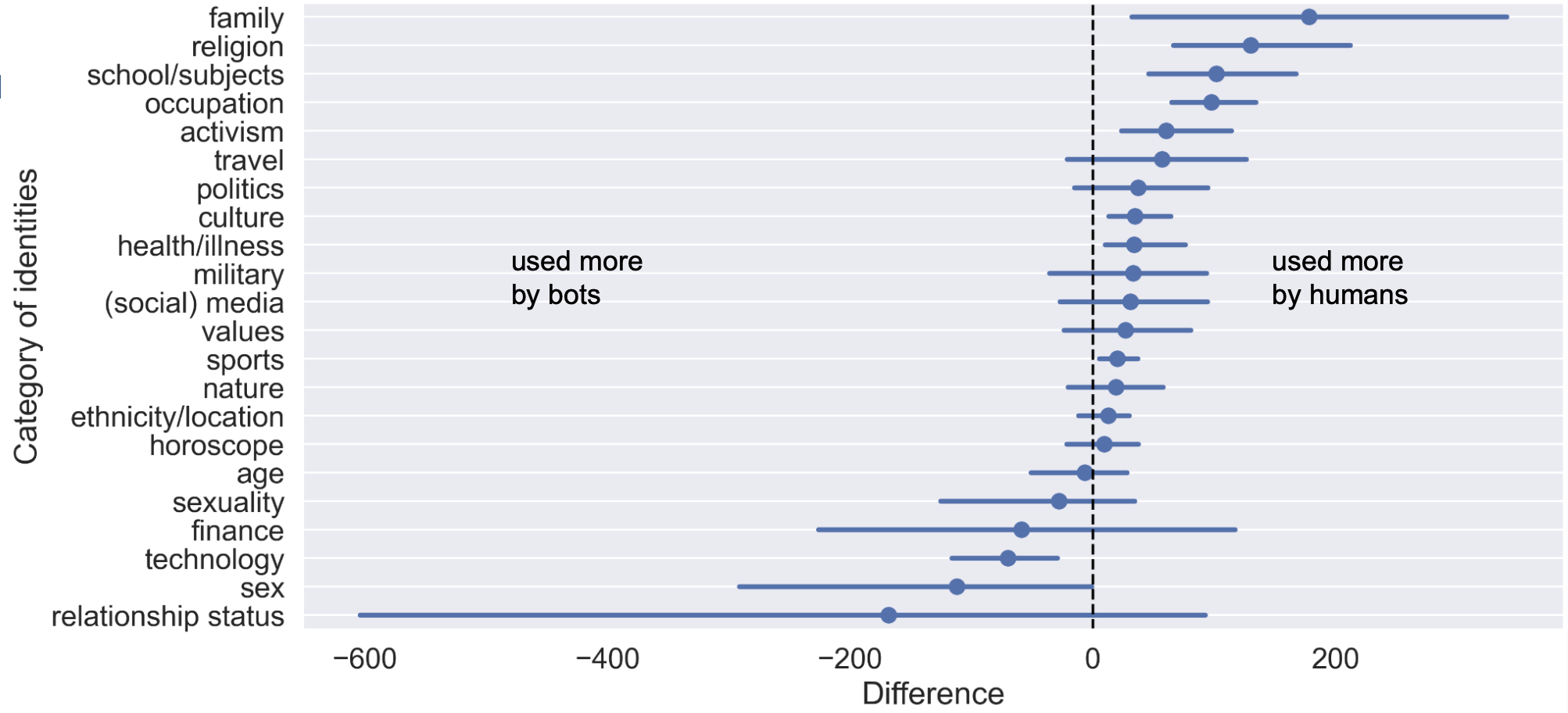}
    \caption{Mean and 95\% CI difference in prevalence per 1 million users for 23 types of identities.  Lower (negative) values are used more by bots, while higher (positive) values are used more by human users.}
    \label{fig:identities-difference}
\end{figure*}

The difference in prevalence per 1 million users of each category of identities in bot and human user profiles is presented in Figure \ref{fig:identities-difference}.
Human users tend to include family, religion, school/subjects, occupation, social activism, culture, and health/illness identifiers more than bots.  
On the other hand, bots include more relationship status, sex, technology, and finance identifiers in their bio than human users.  

Rather than attempt to describe a ``human'' behind the screen, bots sometimes present the specific purpose of their account.
For example, bots have been documented spreading financial (dis)information, which is reflected in the disproportionate use of finance identities like ``markets'', ``e-commerce'', ``investing'', and ``bitcoin'' in bot profiles \cite{tardelli2020characterizing}.  
Moreover, bots are used to spread pornographic content that may or may not be fake\footnote{\url{https://futurefive.co.nz/story/myspace-twitter-targets-for-adult-spam}} with identities in their bio like ``porn'', ``horny'', ``nsfw'', and ``18+ only''.
Many of the technology identities in bot profiles relate to gaming: ``gaming'', ``call of duty'', ``minecraft''.
Notably, bots also include certain signals of relationship status more than humans, such as ``taken'', ``in love with'', and ``happily taken'', which may be an attempt to adopt human-like features.
In sum, approximately 40\% of the 23 categories of identities considered here clearly distinguish bot and human users.  
This confirms (certain types of) identities in bios can be a useful approach to differentiating between human and bot accounts on social media.

\subsection{Comparison of gender identities in bios of bot and human users (RQ2)}

\begin{figure}
    \centering
    \includegraphics[width=\columnwidth]{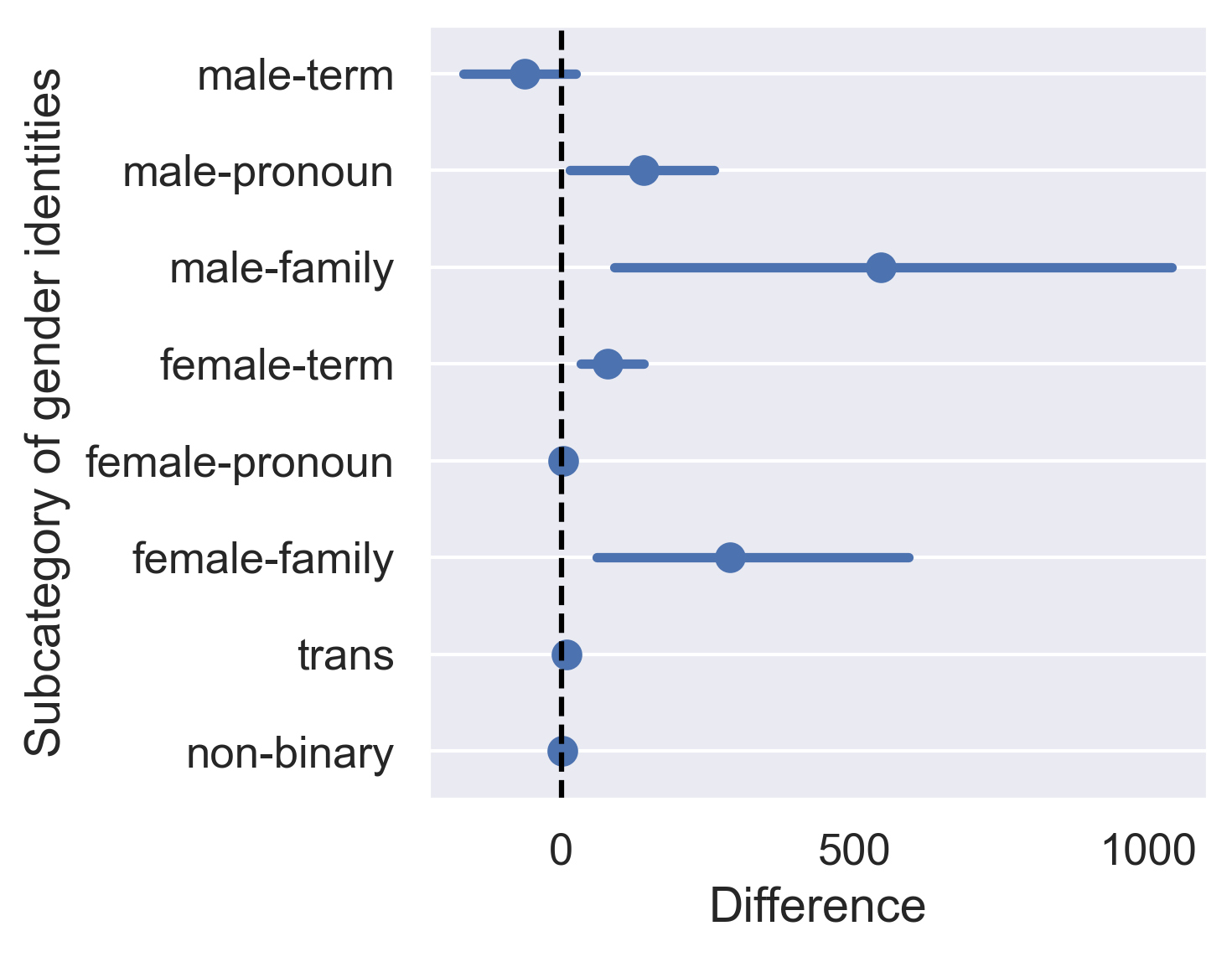}
    \caption{Mean and 95\% CI difference in prevalence per 1M for gender identities.  Lower (negative) values are used more by bots, while higher (positive) values are used more by human users.}
    \label{fig:gender-difference}
\end{figure}

Figure \ref{fig:gender-difference} indicates the difference in prevalence in gender identities between the user types.
As expected given RQ1 results, human accounts use far more male and female family identities than bots.  
Interestingly, female terms like ``girl'', ``woman'' and ``lady'' are used more by human users, while male terms like ``male'', ``boy'', ``man'', and ``mr'' are used far more by bots.
However, male pronouns are used more by human accounts and female pronouns are included at similar rates for both types of users.
Non-binary and transgender identities also are largely not distinguished between user types, however, they also are very scarce in our dataset overall.

Table \ref{tab:gender_user_type} contains the top used gender identities (in order) for bot and human accounts.
The order of use is not wildly different for each user type, with the exception of male terms included more by bots than human users.
This may reflect that bots have adapted to present themselves similarly to human accounts somewhat, although they are less likely to include gender signals than humans.

\begin{table}[]
    \centering
    \begin{tabular}{lp{5.5cm}}
    \hline
    User type & Top 20 gender identities \\
    \hline 
    Bot & girl, man, wife, guy, mother, father, mom, husband, boy, he, her, she, his, dad, mr, him, mommy, woman, daughter, son
 \\
 \hline 
    Human & girl, wife, father, mom, husband, man, mother, guy, dad, he, her, his, she, boy, daughter, woman, him, son, mr, mommy \\ \hline 
    \end{tabular}
    \caption{Top gender identities used by user type}
    \label{tab:gender_user_type}
\end{table}

\subsection{Comparison of gender bias in bios of bot and human users (RQ3)}

\begin{figure}
    \centering
    \includegraphics[width=\columnwidth]{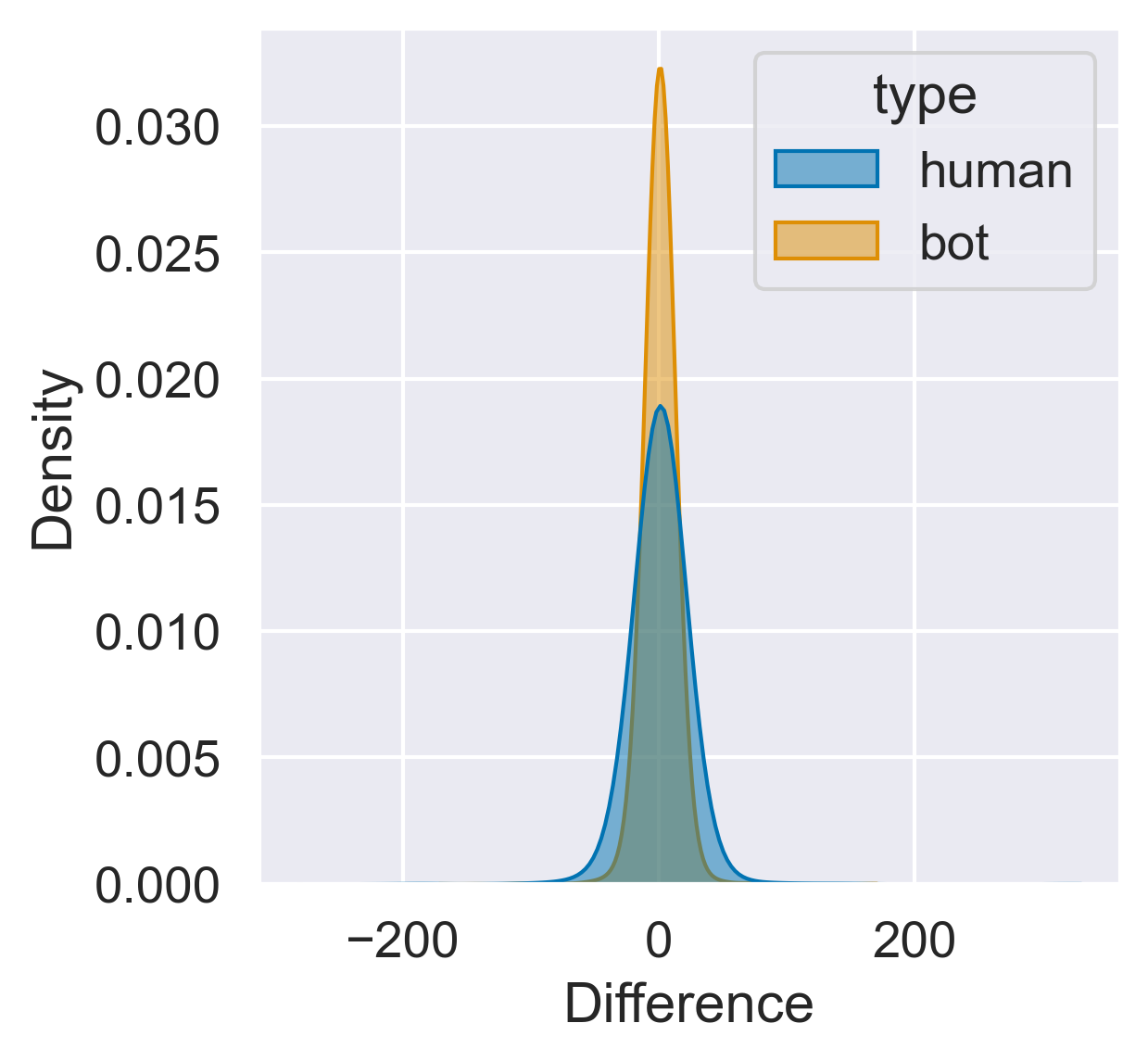}
    \caption{Distribution of differences in weighted log-odds with informative Dirichlet priors between male and female identities for bot and human accounts.}
    \label{fig:fig4}
\end{figure}

The distribution of differences in weighted log-odds along the gender dimension are in Figure \ref{fig:fig4}.  Most notably, the distribution for identities in human profiles is wider than the distribution for bots.  In other words, we find stronger gender bias in human accounts than bot accounts, contrary to our initial assumptions.

\begin{table*}[]
    \centering
    \begin{tabular}{lp{1cm}p{2cm}p{2cm}p{2cm}p{1cm}}
    \hline
    Category\^ & N & Median (H) & Median (B) & T-statistic & M/F\\
    \hline 
    Family & 46 & 10.7 & 6.2 & 3.21** &  F  \\
    \hline 
    Business/finance & 34 & 8.8 & 5.5 & 2.65** &   M \\
    \hline 
    Military & 17 & 8.1 & 3.4 & 1.87  & M  \\
    \hline 
    Technology/cars/gaming & 170 & 7.8 & 5.1 & 4.23***  &  M \\
    \hline 
     Occupation/industry & 454 & 7.2 & 4.1 &  6.84*** &  M \\
    \hline 
     Sports & 424 & 7.0 & 3.9 & 7.81***  &  M \\
    \hline 
    Sexuality & 12 & 6.7 & 2.7 &  1.03  & F \\
    \hline 
    Relationship status & 17 & 6.5 & 3.7 & 1.35 &  M  \\
    \hline 
    Religion & 104 &  6.4 & 3.8 & 3.23** &  M \\
    \hline 
    Nature/animals & 125 & 5.9 & 3.3 &  3.89**  &  M \\
    \hline 
    Health & 31 & 5.7 & 2.8 &  2.39* &  F \\
    \hline 
     Culture & 646  & 5.6 & 3.4 &  7.64*** & F \\
    \hline 
     Horoscope & 26 & 5.4 & 3.3 &  2.82**  &  F \\
    \hline 
    Politics & 60 & 4.9 & 3.1 &  2.05*  &  M\\
    \hline 
    Activism & 63 & 4.4 & 2.8 & 2.3*  & F  \\
    \hline 
    School/subjects & 231 & 3.6 & 2.1 &  4.74***  & M \\
    \hline 
    (Social) media & 62 & 3.6 & 2.3 & 1.34  &  F \\
    \hline 
     Travel & 21 & 3.2 & 1.8 &  1.54  &  F\\
    \hline 
    Race/ethnicity/location & 373 & 3.1 & 1.9 &  6.17***  & M \\
    \hline 
    Values & 17 & 2.8 & 2.1 &  0.91  &  M\\
    \hline 
    Sex & 19 & 2.8 & 2.6 & 0.82  & F   \\
    \hline
     Age & 42 & 2.2 & 1.1 &  2.4*  & F \\
    \hline 
    \end{tabular}
    \caption{\^{} Categories excluding gender identities.  N is the number of identifiers in that category (excl. gender identities), median (H) is the median absolute value of differences in co-occurrence with male and female signals for human accounts, median (B) is the same for bot accounts, t-statistic is from a t-test comparing the strengths of associations with gender signals in human and bot accounts for that category, and M/F represents if identifiers in that category are used with male (M) or female (F) identities more on average. * signifies the significance at $<0.05$, ** signifies significance at $<0.01$, *** signifies significance at $<0.0001$.}
    \label{tab:tab3}
\end{table*}

Next, we consider the average absolute value of the weighted difference in log-odds between male and female edges (``strength of association'') for each type of identity in Table \ref{tab:tab3}.  
This provides a sense of the extent to which different types of identities are divided in their use by gender--- and how it differs for human and bot accounts.
Of the 23 types of identifiers, 15 showed significant differences in association strengths between types of accounts ($\alpha<0.05$).
These categories include family, finance, technology, occupations, and culture.
The remaining 8 categories, such as sexuality and travel, were used to similar extents with both genders across human and bot accounts.
Therefore, while gender bias is more prevalent in human accounts than bot accounts overall, there is variation in the extent to which they differ depending on the type of identifier.

For bots, the types of identifiers that are most biased by gender are family, finance, technology, occupations, sports, and religion.
On the other hand, bot accounts with both male and female gender signals include identifiers that reference age, travel, race/ethnicity/location, school/subjects, and values in their profile to a similar extent.
Echoing previous results, gender bias in bot accounts is not consistent across types of identifiers and requires analysis of specific patterns of associations.

\subsection{Examples of gender bias in bot profiles (RQ4)}

Gender role stereotypes can be expressed in a wide range of forms, from occupations to cultural preferences.  In this section, we delve into the identifiers that co-occur the most with gender signals in bot accounts.  
The identifiers used the most in profiles with male identities (relative to female identities) are``sports'', ``pastor'', ``musician'', ``golfer'', ``engineer'', ``football'', ``gamer'', ``beer'', ``dj'', ``car'', and ``god''.
More broadly, identifiers that relate to business/finance, technology/cars/gaming, and sports are among the most used alongside male identities of bot profiles.

On the other hand, the identifiers used the most in profiles that contain female signals are ``nurse'', ``read'', ``blogger'', ``beauty'', ``fashion'', ``makeup'', ``actress'', ``teacher'', ``cat'', and ``crafter''.
The types of identifiers that co-occur with female signals in bot profiles the most relate to health/illness, travel, horoscope, and relationship status.

\section{Discussion}
Nefarious social bots are a social cyber-security threat \cite{carley2020social} with the demonstrated ability to spread harmful narratives \cite{chang2021social} and facilitate fraud \cite{nizzoli2020charting}.
In this work, we describe and empirically demonstrate a strategy for social bots to mimic human users in an effort to effectively influence other users.
Often, bots are designed to amplify and disseminate messages quickly and efficiently, rather than slowly infiltrate communities and build credibility over time \cite{jacobs2023tracking}.
Many bots do not have any bio at all, hence we do not have an estimate for this value.
Here, we focus on bots that include identities in their profile, possibly to increase human-like qualities and better blend in with a social network.

We use self-designated labels of binary gender (e.g., pronouns, family roles) to examine existing identifier use and gender biases in X user bios, where identifiers include social identities, preferences, activities, and affiliations.
Using self-designated labels provides the indication of the gender that people affiliate themselves with, rather than a fully inferred gender type. 

Our findings reveal some types of identifiers are used more in profiles of bots than human accounts, such as sex, technology, and finance.
This is especially relevant given the recent influx of ``porn bots'' on X\footnote{\url{https://www.rollingstone.com/culture/culture-features/twitter-porn-bot-sex-meme-1234973030/}}.
Profiles with these types of identities, along with other attributes, may be among the most recognizable by humans.
Furthermore, bots often use identities to refer to the purpose of the account rather than a ``human'' behind the account.

Overall, bots do not use gender signals in their bio at the same rate as human users. Moreover, humans use more identifiers that are strongly associated with one gender or the other than bots.  This contradicts our initial expectation that bots will use exaggerated gender role stereotypes in an effort to invoke familiarity biases.
However, we are still able to find specific examples where bot profiles are biased by gender.
The most commonly co-occurring identifiers with female identities in bot bios relate to beauty, art, and care-taking occupations like nursing and teaching.
Conversely, the most commonly co-occurring identifiers with male identities are about sports, gaming, beer, and religion. The occupations used with male signals in bot profiles the most are in music or technical fields.

An aim of this paper is to encourage other researchers to further investigate the ways social identities (e.g., gender, age, religion) can be leveraged by malicious actors to increase manipulation ability, especially in anonymous online settings.  
It is also relevant to those studying AI-generated content online, as many of the most popular models are rife with biases \cite{kotek2023gender}.
The content generated by these models reflect the biases in the training data, which is presumably mostly human generated.
Identifying patterns of associations perpetuated by LLMs and other forms of generative models may prove helpful in identifying AI-generated content from not.


\paragraph{Designing Constructive Bot Detection Guides}
Social media platforms are conduits for individuals to share their thoughts, however, this comes at a social cyber-security risk \citep{carley2020social}. In the digital world, where anonymity and false identities are present, the analysis of the affiliation of users and their behavior are important to differentiate truth from false. 
It has been well-established that humans are not very good at distinguishing between human and bot users \cite{kenny2024duped} or AI- and human-generated (brief) written self-presentations \cite{jakesch2023human}. 
This research contributes towards intelligent detection of deceptive entities (bots) on social media platforms. Through understanding the self-presentation of bot and human identities, analysts can identify categories of identities that clearly differentiate between both user types, like the ones provided here. 
A multitude of tools have been developed to detect bots using certain characteristics such as the rate of posting, follower/following ratio and so forth. Examples of such algorithms are Botometer and BotBuster \cite{ng2023botbuster,yang2022botometer}. 
Our analysis can be used to inform future bot detection algorithms, where detection models can be tuned to better leverage patterns of gender stereotypes to classify users as bots and humans.

In addition, there is evidence that people can improve their bot detection ability through training \cite{kenny2023improving}.
These trainings can be supplemented to include explanations of how bots leverage human biases (e.g., confirmation bias) to gain influence, similar to existing misinformation inoculation materials \cite{lewandowsky2021countering}.
Although many users who use gender signals are indeed humans, profiles including gender role stereotypes may warrant more consideration before being trusted, especially if they relate to technology or business/finance.

\paragraph{Limitations and Future Work}
Gendered identities evolve continually, and therefore the terms and identities that we used may be outdated with the times \cite{levitt2019psychosocial}. 
Non-binary and transgender identities were very rare in this dataset ($<$0.01\% of bios) which may reflect the prevailing gender trends of the time period in which the data is collected; or the gaps in our extraction and/or labelling process; or the trend of the inclusion of gender identities.
Future work involves exploring more divisions of gender identities, especially beefing up the set of terms that correspond to non-binary or transgender identities within social media lingo.

Further, we do not attempt to classify the gender user profile names or images.
Name-based gender detection is not straightforward, for a name can be representative as a female or male names \citep{van2023open,lockhart2023name}.
While gender detection in social media bios can be improved through image-based gender detection \citep{schwemmer2020diagnosing}, studies have shown that there are mismatches of gender from the profile picture and the names, especially in bot accounts \citep{yang2024characteristics}. 
The evolving definitions of gender \citep{fraser2015gender} and the displacement of gender between text and image makes gender detection in social media still an open research question.
Future work should compare gender identities presented through names, bios, and images.
Research shows a bio is not always necessary for a bot to become influential \cite{aiello2012people}.
Future work should also empirically examine the extent of, if any, advantage bots have by customizing their bio to include common stereotypes.

Finally, the validity of the comparisons in this work hinge on the correctness of BotBuster, the bot detection tool used \cite{ng2023botbuster}.
BotBuster was trained on several manually annotated datasets with bot and human labels, similar to other bot detection algorithms like Botometer \cite{varol2017online} and BotHunter \cite{beskow2018bot}. 
User features, such as user bios, are used by these bot detection algorithms to differentiate between human and bot users. 
These algorithms consist of machine learning models, which are generally black boxes. 
Recent work performed a feature analysis study on BotBuster, showing that there are groups of terms in bios that are more associated with bot accounts \cite{ng2024assembling}. 
These biases are perpetuated in classifications, biasing our analysis of differences in identity use in bot and human profiles to some extent. 
Nonetheless, identities in bios is just one of many features used to differentiate bot and human users by BotBuster.
Future work should compare differences in identities used by bot and human accounts that are detected by other bot detection tools with our results (e.g., \citet{varol2017online,beskow2018bot}).

\subsubsection{Ethical statement}
The purpose of this work is to improve understanding on how covert bots leverage gender biases in their profile. We do not attempt to examine actual gender differences in specific identities. While the work can inform researchers building bot detection algorithms, this work can also be used by nefarious bot operators to avoid detection and build workarounds. Nonetheless, it is still important to understand the existence of differences of identities between bots and humans as a research problem, such that it can be continually studied and used as indicators for the evolution of bots and the need for algorithm refresh.

\bibliography{biblio}

\subsubsection{Acknowledgments.}
This work was supported in part by the Knight Foundation and the Office of Naval Research grant Minerva-Multi-Level Models of Covert Online Information Campaigns, N000142112765, N000141812106, and N000141812108.
Additional support was provided by the center for Informed Democracy and Social-cybersecurity  (IDeaS) and the center for Computational Analysis of Social and Organizational Systems (CASOS) at Carnegie Mellon University. 
The views and conclusions contained in this document are those of the authors and should not be interpreted as representing the official policies, either expressed or implied, of the Knight Foundation,  Office of Naval Research, or the U.S. Government.

\newpage 

\appendix

\onecolumn

\begin{longtable}{p{.15\textwidth}p{.75\textwidth}}
    Identifier type & Identifiers \\
    \hline
(social) media	& ['news', 'social media specialist', 'twitch streamer', 'vlogger', 'food blogger', 'content strategist', 'pjnet', 'current affairs', 'content creator', 'broadcast journalist', 'tyler oakley', 'youtube addict', 'blogger', 'tweeter', 'journalism', 'multimedia journalist', 'bbc', 'retweeter', 'social media addict', 'kian lawley', 'social media expert', 'troll', 'social media manager', 'digitalmarketing', 'wsj', 'dolan twins', 'cameron dallas', 'streamer', 'social media marketer', 'youtube content creator', 'gab', 'watching youtube', 'blog', 'columbiajourn', 'columnist', 'media', 'politico', 'youtube gamer', 'social media guru', 'content marketing', 'social media consultant', 'youtuber', 'social media junkie', 'youtubers', 'social media enthusiast', 'jacobsartorius', 'joe sugg', 'reporter', 'small youtuber', 'social media strategist', 'occasional blogger', 'broadcaster', 'journalist', 'influencer', 'make youtube videos', 'npr', 'cnn', 'amazingphil', 'camerondallas', 'blogging', 'nytimes', 'tgdn']
       \\
    \hline
activism &	['community activist', 'revolutionary', 'rotarian', 'pacifist', 'unity', 'nationalist', 'intersectional feminist', 'community manager', 'advocate', 'community builder', 'woke', 'loveislove', 'humanist', 'environmentalist', 'social issues', 'feminism', 'justice', 'environmental', 'ally', 'conservationist', 'freepalestine', 'humanitarian', 'jft96', 'activist', 'community organizer', 'feminist', 'allblacklivesmatter', 'blacklivesmatter', '2a', 'alllivesmatter', 'nodapl', 'world changer', 'social justice', 'nra', 'prolife', 'concerned citizen', 'advocacy', 'theresistance', 'guns', 'socialjustice', 'freemason', 'equity', 'philanthropist', 'community leader', 'repealthe8th', 'sustainability', 'humanrights', 'activism', 'community volunteer', 'social activist', 'bluelivesmatter', 'climate change', 'volunteering', 'giver', 'change agent', 'conservation', '1a', 'volunteer', 'human rights', 'blm', 'lovewins', 'resistance', 'tgdn'] \\
    \hline
age &	['class of 2018', '23 years old', "o '17", '17 years old', '21 years old', 'adult', '19 y', '11 years old', '13 years old', '18 y', 'gatornation', '18 years old', 'millennial', 'high school', 'teen', '27 years old', 'young', '8th grade', '25 years old', 'teenager', '16 years old', 'class of 2017', '13 years young', 'o 2015', "class of '18", '20 years old', 'sophomore', 'grown', '22 years old', 'o 2016', '12 years old', 'old', 'class of 2016', 'freshman', '24 years old', '19 years old', '26 years old', '15 years old', "90's baby", '14 years old', '16 years young', 'senior'] \\
    \hline
business/finance &	['financial', 'economy', 'e-commerce', 'investing', 'brokerage', 'management', 'business', 'starbucks', 'entrepreneurship', 'entrepreneurial', 'fintech', 'startup', 'founder', 'startups', 'nike', 'international business', 'stock trader', 'markets', 'bitcoin', 'trading', 'investment', 'mlm', 'finance', 'sony', 'stocks', 'investments', 'investor', 'marketing', 'forex trader', 'company', 'billionaire', 'microsoft', 'ibm', 'corporate'] \\
    \hline
culture	& ['cheese lover', 'weeknd', 'cinephile', 'dj', 'caterer', 'loves to cook', 'justin bieber', 'theweeknd', 'remixer', 'chris smith', 'bibliophile', 'foody', 'louis tomlinson', 'performing', 'quilter', 'supergirl', 'kdrama', 'u2', 'shawn mendes', 'john cena', 'shopaholic', 'comic books', 'voice', 'food eater', 'rihanna', 'dexter', 'tattoos', 'sketching', 'twenty one pilots', 'curry', 'bap', 'gaga', '9teen', 'ed sheeran', 'bathroom singer', 'sci fi', 'playwright', 'zayn malik', 'poker player', 'artists', 'ofwgkta', 'mumbaikar', 'food blogger', 'country music', 'house music', 'cinematographer', 'picture taker', 'food lover', 'shoe lover', 'singers', 'wonder woman', 'anglophile', 'freelance writer', 'walking dead', 'ahs', 'dubstep', 'reggae', 'literature', 'taking photos', 'harry styles', 'tv junkie', 'ballet', 'troye sivan', 'spongebob', 'renaissance man', 'nct', 'gypsy', 'archivist', 'portrait photographer', 'cosplay', 'writer', 'alt', 'zumba', 'bts', 'dancehall', 'musicals', 'twilight', 'debate', 'make stuff', 'one direction', 'janoskians', 'poems', 'blues', 'fruit', 'guitarist', 'craft beers', 'hobbit', 'story teller', 'writter', 'booklover', 'cinematography', 'transformers', 'crocheter', 'composer', 'divergent', 'arianagrande', 'watch tv', 'knitter', 'chess', 'michael clifford', 'fifth harmony', 'drums', 'disney', 'pizza enthusiast', 'design', 'write', 'lyricist', 'sound designer', 'ymcmb', 'song-writer', 'multi-instrumentalist', 'beingsalmankhan', 'bookworm', 'beatboxer', 'bieber', 'photog', 'retro', 'instrumentalist', 'illustration', 'actress', 'directioner', 'films', 'songwriters', 'pianist', 'sew', 'classic rock', 'riverdale', 'wannabe writer', 'photo editor', 'song writter', 'radio', 'criminal minds', 'loves reading', 'wordsmith', 'at the disco', 'marching band', 'visual designer', 'flash', 'cooking', 'crafts', 'audiophile', 'podcast host', 'ikon', 'crochet', 'author', 'vocalist', 'videographer', 'tv host', 'pop', 'pll', 'superman', 'pizza lover', 'pastry chef', 'raver', 'rockstar', 'superhero', 'shakira', 'new music', 'anime', 'reiki master', 'indie', 'upcoming artist', 'choreographer', 'arianator', 'weeb', 'loves cooking', 'photojournalist', 'alternative', 'artwork', 'survivor', 'dancing', 'srk', "dylan o'brien", 'punk', 'selenagomez', 'graphics designer', 'play bass', 'amwriting', 'cupcake', 'exo', 'always hungry', 'gluten free', 'entertainment', 'record producer', 'harmonizer', 'bollywood', 'ukulele', 'breaking bad', 'painting', 'gotham', 'd\&d', 'cartoonist', 'big brother', 'drinker', 'minimalist', 'tv watcher', 'novels', 'roman reigns', 'screenwriter', 'superheroes', 'tea enthusiast', 'drunk', 'boybands', 'artist', 'create', 'nani', 'sherlockian', 'taking pictures', 'craft beer enthusiast', 'starbucks', 'scotch', 'art', 'beat maker', 'make music', 'seuss', 'books', 'piano', 'performing arts', 'paint', 'steampunk', 'interior designer', 'tea addict', 'comic', 'bbq', 'loves music', 'jack johnson', 'pokemon', 'beets', 'debater', 'copywriter', 'bourbon', 'watch movies', 'larry stylinson', 'performer', 'starwars', 'sherlock', 'paramore', 'graphic design', 'cameraman', 'kanye west', 'khan', 'making videos', 'funk', 'cinema', 'hamilton', 'cigars', 'otaku', 'musicislife', 'kiwi', 'coco chanel', '3d artist', 'trancefamily', 'animation', 'percussionist', 'knit', 'ladygaga', 'beer', 'tv producer', 'veggie', 'podcasts', 'comedian', 'bmth', 'tv addict', 'magcon', 'occasional writer', 'performing artist', 'cartoons', 'craft beer lover', 'song writer', 'arctic monkeys', 'read', 'vagabond', 'tv presenter', 'seamstress', 'vocals', '1975', 'spnfamily', 'captain of my soul', 'rattpack', 'classics', 'cold beer', 'illustrator', 'harry styles', 'potterhead', 'edm', 'paleo', 'gleek', 'r\&b', 'halsey', 'coldplay', '1d', 'reading', 'j cole', 'graphic', 'amateur chef', 'designer', 'taylor swift', 'k-pop', 'eater', 'editing', 'aspiring artist', 'sneakers', 'cole world', 'bands', 'die hard', 'play guitar', 'band obsessed', 'thespian', 'artiste', 'wino', 'justinbieber', 'metallica', 'brewer', 'little mix', 'taco bell', 'loves photography', 'kurt cobain', 'sneaker head', 'drinking', 'novelist', 'techno', 'good book', 'ariana grande', 'sketch', 'woodworker', 'marvel', 'play music', 'dancer', 'fashionista', 'amateur photographer', 'chocolate addict', 'interior design', 'harrypotter', 'skinhubsquad', 'rap', 'shawol', 'image consultant', 'dance', 'documentary filmmaker', 'cartoon', 'mgk', 'rapper', 'zendaya', 'chocoholic', 'musiclover', 'deadpool', 'taylorswift13', 'drawer', 'teenwolf', 'keyboardist', '2ne1', 'photography', 'coleworld', 'drink beer', 'peter pan', 'good music', 'lana del rey', 'camera', 'drink coffee', 'fashion', 'loves to sing', 'storyteller', 'sag', 'violinist', 'good beer', 'improv', 'sneakerhead', 'drum', 'sam smith', 'heavy metal', 'music', 'katy perry', 'favorite movie', 'bruno mars', 'dancers', 'outlander', 'animator', 'rocker', 'tattoo', 'salsa', 'voracious reader', 'drawing', 'musician', 'disneyland', 'guitar', 'trekkie', 'potter', 'musical', 'avenged sevenfold', 'swiftie', 'snsd', 'francophile', 'zquad', 'graphicdesigner', 'selena gomez', 'cook', 'homebrewer', 'video', 'lotr', 'lovatic', 'doctorwho', 'salman khan', 'melanie martinez', 'percussion', 'voiceover', 'disco', 'demi lovato', 'gameofthrones', 'belieber', 'spoken word artist', 'whovian', 'oitnb', 'published author', 'teen wolf', 'simba', 'make beats', 'watching movies', 'staff writer', 'baking', 'movies', 'mark twain', 'linkin park', 'liam payne', 'diehard', 'soprano', 'reader', 'union j', 'podcaster', 'poet', 'melody', 'hip hop'] \\
\hline
culture cont. & ['oasis', 'tv fanatic', 'diyer', 'visual artist', 'harry potter', 'elvis', 'arts', 'basshead', 'sci-fi', 'shakespeare', 'frozen', 'trumpet', 'comics', 'violin', 'eminem', 'tattoo artist', 'cookie monster', 'take photos', 'rock', 'vampire diaries', 'movie', 'ncis', 'katyperry', 'vocal coach', 'loves anime', 'script writer', 'enjoy reading', 'cosplayer', 'cheesecake', 'americana', 'beatles', 'stand up comedian', 'poetry', 'game of thrones', 'film', 'aspiring author', 'star trek', 'recording artist', 'filmmaking', 'salman', 'red velvet', 'tyga', "grey's anatomy", 'gossip girl', 'culture', 'brony', 'stage manager', 'artisan', 'food addict', 'theater', 'book', 'hard rock', 'songwriter', 'reality tv junkie', 'electro', 'tolkien', 'choir', 'manga', 'food enthusiast', 'batman', '5 seconds of summer', 'baker', 'imagine dragons', 'saxophonist', 'chef', 'foodie', 'hunger games', 'band geek', 'bring me the horizon', 'painter', 'suits', 'meditator', 'jordy', 'make videos', 'rhiannon', 'pink floyd', 'filmmaker', 'acting', 'vfx', 'songs', 'opera singer', 'food junkie', 'entertainer', 'drummer', 'watching tv', 'television', 'tea lover', 'onedirection', 'jedi', 'drama', 'tv personality', 'barista', 'whiskey', 'battlestar galactica', 'play the guitar', 'niallofficial', 'coffee', 'pizza expert', 'real liam payne', 'philadelphia', 'fiction', 'beyhive', 'iamsrk', 'costume designer', 'metal', 'selenator', 'foodlover', 'eat', 'bake', 'shadowhunters', 'comedy', 'scrapbooker', 'adele', 'ska', 'fall out boy', 'sound engineer', 'videos', 'singer', 'michael jackson', 'miley cyrus', 'guitars', 'harley quinn', 'doodler', 'chocolate lover', 'cellist', 'actor', 'beyonce', 'craftbeer', 'play piano', 'scriptwriter', 'trance', 'mcfly', 'nickiminaj', 'star wars', 'horror', '5sos', 'diet', 'moonchild', 'book lover', 'monsta x', 'concept artist', 'kpop', 'r5', 'drawings', 'nutrition', 'troye', 'food fanatic', 'fusion', 'technical writer', 'master of none', 'improviser', 'bass', 'acoustic', 'mcr', 'play drums', 'tea drinker', 'blondie', 'copy editor', 'grunge', 'watching anime', 'teambreezy', 'hip-hop', 'thewalkingdead', 'jetlife', 'flute', 'metalhead', 'home brewer', 'got7', 'singing', 'beef', 'graphic designer', 'hiphop', 'draw', 'uke', 'broadway', 'sculptor', 'scifi', 'blackpink', 'nirvana', 'visual arts', 'brewers', 'glee', 'making music', 'pineapples', 'aspiring writer', 'recording engineer', 'chris brown', 'crafter', 'take pictures', 'concert goer', 'writing', 'nash grier', 'trap', 'vegetarian', 'superwoman', 'cake decorator', 'bassist', 'photographer', 'nicki minaj', 'listen to music', 'avid reader', 'demi', 'theatre', 'make art', 'fine artist', 'listening to music', 'igot7', 'wine', 'jazz', 'niall horan', 'beatmaker', 'lead singer', 'vegan', 'home cook', 'amateur writer', 'illustrations', 'sing'] \\
\hline
disability/(mental) illness	& ['adhd', 'diabetic', 'ocd', 'hiv', 'addict', 'suicidal', 'diabetes', 'disabled', 'disability', 'anxiety', 'breast cancer survivor', 'depressed', 'autism', 'autistic', 't1d', 'self harm', 'nfb', 'depression', 'ptsd', 'type 1 diabetic', 'mental health', 'fuckcancer', 'deaf', 'caffeine addict', 'spoonie', 'bipolar', 'alcoholic', 'fibromyalgia', 'addiction', 'mentalhealth', 'sober'] \\
    \hline
family & 	['full time mom', 'sister', 'thefamily', 'my wife', 'mummy', 'parent', 'nana', 'homemaker', 'aunt', 'family', 'momma', 'mom', 'papa', 'lucky husband', 'stepmom', 'genealogy', 'loves family', 'milf', 'spouse', 'full time mommy', 'father', 'devoted husband', 'uncle', 'happily married', 'proudfather', 'supermom', 'grandchildren', 'cousin', 'brother', 'mompreneur', 'widow', 'grandma', 'granddaughter', 'two kids', 'fiance', 'godfather', 'housewife', 'grandfather', 'niece', 'mum', 'savior', 'my daughter', 'mother', 'devoted wife', '4 kids', 'twins', 'newborn', 'my kids', 'happy husband', 'parenting', 'son', 'married', 'teammommy', 'stay at home mom', 'new dad', 'house wife', 'twin', 'aunty', 'dad', 'sahm', 'grandad', 'my children', 'my husband', 'parents', 'woman', 'super mom', 'grandson', 'grandmother', 'mommy', '3 kids', 'big sister', '2 boys', 'being a mom', 'hubby', '2 kids', 'rip dad', 'sis', 'mama', 'grandkids', 'grandparent', 'bro', 'wife', 'nephew', 'my family', 'my son', 'full time dad', 'children', 'motherhood', 'grandpa', 'no kids', 'my hubby', 'spending time with family', 'new mom', 'spending time with my family', 'wifey', 'happy wife', 'granny', 'daughter', 'auntie', 'grammy', 'familyfirst', 'daddy', 'brothers', 'husband'] \\
    \hline
horoscope	&['sagittarius', 'virgo', 'pisces', 'astrology', 'teamsagittarius', 'scorpio', 'teampisces', 'teamcancer', 'teamgemini', 'aquarian', 'libra', 'aquarius', 'cancerian', 'teamtaurus', 'teamlibra', 'leo', 'cancer', 'astronomy', 'taurus', 'aries', 'teamvirgo', 'teamaries', 'capricorn', 'teamscorpio', 'teamleo', 'gemini'] \\
    \hline
military	& ['usarmy', 'marine', 'usaf', 'usn', 'gunner', 'navy', 'usmc', 'soldier', 'sniper', 'air force', 'army', 'armystrong', 'military', 'semper fi', 'veteran', 'veterans', 'former marine'] \\
    \hline
location/ethnicity/ race	& ['cornwall', 'rochester', 'south africa', 'new yorker', 'birmingham', 'ga', 'global citizen', 'pakistan', 'minneapolis', 'beach bum', 'nc', 'punjabi', 'korean', 'tulsa', 'memphis', 'merica', 'urbanist', 'louisville', 'us', 'la', 'au', 'yeg', 'puerto rican', 'african', 'montana', 'new mexico', 'minnesota', 'boston', 'northern ireland', 'newyork', 'nm', 'southampton', 'richmond', 'lebanese', 'ghanaian', 'lax', 'dmv', 'atx', 'malaysia', 'indy', 'kenyan', 'southern', 'portland', 'rva', 'france', 'glasgow', 'va', 'vietnam', 'denmark', 'english', 'croatia', 'milwaukee', 'arkansas', 'small town', 'pennsylvania', 'beaches', 'chennai', 'winnipeg', 'uk', 'ireland', 'u k', 'nola', 'york', 'greece', 'mexican', 'essex', 'puertorican', 'nj', 'cambridge', 'punjab', 'bali', 'detroit', 'filipino', 'sheffield', 'philly', 'miami', 'seattle', 'spanish', 'indonesia', 'turkey', 'palestine', 'hispanic', 'san francisco', 'florida', 'indiana', 'cardiff', 'foreign', 'aussie', 'sd', 'melanin', 'paris', 'citizen', 'italy', 'immigration', 'l a', 'malaysian', 'mexico', 'caribbean', 'jamaican', 'iowa', 'kenya', 'middle east', 'china', 'dorset', 'tampa', 'dc', 'u s', 'raleigh', 'nz', 'texan', 's c', 'vancouver', 'atlanta', 'brooklyn', 'chinese', 'michigan', 'german', 'japan', 'california', 'lebanon', 'pa', 'mumbai', 'alaska', 'nv', 'madridista', 'dallas', 'georgetown', 'haitian', 'japanese', 'houston', 'ghana', 'columbus', 'buffalo', 'bangalore', 'trinidad', 'colorado', 'vietnamese', 'north', 'scotland', 'portsmouth', 'bay area', 'russian', 'mixed', 'belfast', 'cape town', 'egyptian', 'freepalestine', 'hong kong', 'dubai', 'stl', 'korea', 'irish', 'chicagoan', 'usmnt', 'americafirst', 'san antonio', 'las vegas', 'brisbane', 'louisiana', 'alexandria', 'toronto', 'manchester', 'alabama', 'lancashire', 'colombian', 'p a', 'sweden', 'europe', 'fl', 'cuban', 'omaha', 'fiji', 'dutch', 'htx', 'delhi', 'polish', 'savannah', 'french', 'nottingham', 'mi', 'egypt', 'indonesian', 'brazilian', 'nyc', 'welsh', 'persian', 'latin', 'arabic', 'los angeles', 'pdx', 'india', 'small town girl', 'de', 'melbourne', 'poland', 'singapore', 'immigrant', 'arizona', 'american', 'cincinnati', 'my india', 'adelaide', 'canada', 'australian', 'based in london', 'ontario', 'dominican', 'asian', 'amsterdam', 'idaho', 'iraq', 'hyderabad', 'nepal', 'east coast', 'nevada', 'nashville', 'jersey', 'bronx', 'philippines', 'liverpool', 'ct', 'maine', 'netherlands', 'wales', 'edinburgh', 'delaware', 'barcelona', 'israel', 'latina', 'latino', 'usa', 'chicago', 'plymouth', 'lagos', 'peruvian', 'ny', 'russia', 'oregon', 'mke', 'new jersey', 'phoenix', 'midwest', 'cleveland', 'nigerian', 'indian', 'finland', 'yorkshire', 'boricua', 'londoner', 'hawaii', 'proudly south african', 'asia', 'utah', 'syrian', 'celtic', 'madrid', 'kashmir', 'living in london', 'washington', 'iran', 'london', 'long island', 'made in the usa', 'west coast', 'aberdeen', 'long walks on the beach', 'tx', 'surrey', 'leicester', 'gbr', 'nebraska', 'uae', 'kent', 'wv', 'mississippi', 'brit', 'kansas', 'jakarta', 'az', 'australia', 'san diego', 'charleston', 'wisconsin', 'blackmafia', 'oxford', 'west ham', 'massachusetts', 'european', 'wi', 'vegas', 'south', 'portuguese', 'filipina', 'kentucky', 'texans', 'country girl', 'fort worth', 'denver', 'africa', 'germany', 'oklahoma', 'jacksonville', 'oakland', 'philadelphia', 'orange county', 'newcastle', 'north carolina', 'swedish', 'eng', 'brighton', 'sf', 'missouri', 'indianapolis', 'bulgaria', 'bradford', 'tucson', 'belgium', 'greek', 'georgia', 'norwich', 'spain', 'kc', 'swansea', 'connecticut', 'texas', 'british', 'england', 'bangladesh', 'uganda', 'ohio', 'new delhi', 'reno', 'new zealand', 'scottish', 'blackgirlmagic', 'jamaica', 'dublin', 'virginia', 'hertfordshire', 'pittsburgh', 'kuwait', 'karachi', 'pakistani', 'new york', 'tennessee', 'island girl', 'beach lover', 'nigeria', 'syria', 'canadian', 'cali', 'orlando', 'switzerland', 'sacramento', 'redneck', 'milan', 'bristol', 'austin', 'america', 'south african', 'baltimore', 'berlin', 'tokyo', 'uswnt', 'ca', 'leeds', 'socal', 'illinois', 'bournemouth', 'managing partner', 'norway', 'new orleans', 'italian', 'palestinian', 'd c', 'lahore', 'maryland'] \\
    \hline
occupation/industry	& ['risk management', 'prof', 'nasa', 'nursing', 'dean', 'game designer', 'dj', 'caterer', 'truck driver', 'magician', 'ambassador', 'events manager', 'exec', 'banker', 'urban planner', 'beekeeper', 'researcher', 'infosec', 'psychiatrist', 'social media specialist', 'salon owner', 'product development', 'communications consultant', 'twitch streamer', 'secretary', 'talent manager', 'first grade teacher', 'it manager', 'playwright', 'ceo', 'counselor', 'poker player', 'boss', 'worship leader', 'logistics', 'psychologist', 'police', 'webdeveloper', 'lifeguard', 'chartered accountant', 'managing director', 'cinematographer', 'e-commerce', 'property manager', 'developer', 'adjunct professor', 'auditor', 'data scientist', 'freelance writer', 'sociologist', 'it engineer', 'pediatrician', 'agent', 'productions', 'hairstylist', 'content strategist', 'proofreader', 'instructional coach', 'security', 'nurse', 'archivist', 'anthropologist', 'portrait photographer', 'writer', 'reviewer', 'ghostwriter', 'specialist', 'promoter', 'lighting designer', 'architecture', 'announcer', 'gambler', 'faculty', 'chiropractic', 'guitarist', 'medic', 'mover', 'composer', 'marketer', 'content creator', 'community manager', '3rd grade teacher', 'cpa', 'rocket scientist', 'broadcast journalist', 'online marketer', 'data analyst', 'freelancer', 'executive director', 'statistician', 'sound designer', 'civil engineer', 'song-writer', 'operations manager', 'web designer', 'lecturer', 'actress', 'pr', 'personal trainer', 'trainer', 'board member', 'chairman', 'entreprenuer', 'plumber', 'visual designer', 'healthcare', 'pe teacher', 'consulting', 'construction', 'cto', 'program manager', 'doc', 'producer', 'event manager', 'social studies teacher', 'author', 'vocalist', 'videographer', 'tv host', 'pastry chef', 'front end developer', 'accountant', 'mompreneur', 'educationist', 'administrator', 'virtual assistant', 'hvac', 'public health', 'engineer', 'scientist', 'pilot', 'printmaker', 'executive', 'graphics designer', 'model', 'serial entrepreneur', 'record producer', 'beauty', 'neuroscientist', 'multimedia journalist', 'cartoonist', 'freelance', 'software developer', 'realestate', 'screenwriter', 'translator', 'handyman', 'builder', 'decorator', 'massage therapist', 'game developer', 'aviator', 'manager', 'social worker', 'interior designer', 'digitalhealth', 'contractor', '2nd grade teacher', 'assistant professor', 'emcee', 'farmer', 'event organizer', 'copywriter', 'account manager', 'educator', 'real estate', 'public relations', 'medical', 'dental', 'performer', 'datascience', 'make up artist', 'florist', 'lawyer', 'cameraman', 'flight attendant', 'hair stylist', 'nanny', 'hairdresser', 'businessman', 'web developer', 'entrepeneur', 'esq', '3d artist', 'cfp', 'tv producer', 'self-employed', 'network engineer', 'comedian', 'performing artist', 'ems', 'pharma', 'structural engineer', 'judge', 'song writer', 'life coach', 'licensed cosmetologist', 'physician', 'tv presenter', 'office manager', 'career coach', 'seamstress', 'founder', 'accounting', 'assistant director', 'motivational speaker', 'pharmacy', 'law', 'co founder', 'illustrator', 'founding member', 'publicity', 'future teacher', 'biologist', 'it consultant', 'stock trader', 'hotelier', 'product management', 'psychotherapist', 'chemist', 'doula', 'librarian', 'historian', 'event coordinator', 'mining', 'dentist', 'herbalist', 'bartender', 'primary school teacher', 'owner', 'inventor', 'jewelry maker', 'data science', 'advisor', 'anchor', 'electrical engineer', 'banking', 'occupational therapist', 'registered dietitian', 'cosmetologist', 'social media marketer', 'doctor', 'image consultant', 'youtube content creator', 'former teacher', 'mathematician', 'documentary filmmaker', 'social entrepreneur', 'architect', 'rapper', 'columnist', 'instructional designer', 'counsellor', 'estate planning', 'curator', 'instructor', 'hr', 'networker', 'violinist', 'aviation', 'farming', 'realtor', 'musician', '5th grade teacher', 'programmer', 'meteorologist', 'teacher', 'graphicdesigner', 'event planner', 'managing editor', 'broker', 'speaker', 'audio engineer', 'carpenter', 'salesman', 'vet', 'ecologist', 'public servant', 'social media consultant', 'ux designer', 'youtuber', 'sales', 'it specialist', 'published author', 'veterinarian', 'hospitality', 'staff writer', 'software engineer', 'paramedic', 'preschool teacher', 'cosmetology', 'genealogist', 'recruiter', 'account executive', 'geographer', 'podcaster', 'waitress', 'chiropractor', 'esthetician', 'dr', 'communications specialist', 'website designer', 'visual artist', 'project manager', 'self employed', 'electrician', 'brand strategist', 'program director', '4th grade teacher', 'diplomat', 'socialite', 'registered nurse', 'tattoo artist', 'co-owner', 'officer', 'makeup', 'criminal justice', 'retired', 'webdesigner', 'businesswoman', 'co-founder', 'script writer', 'make-up artist', 'aspiring author', 'landlord', 'co-host', 'assistant principal', 'reporter', 'stage manager', 'archaeologist', 'unemployed', 'head coach', 'salesforce', 'songwriter', 'project management', 'scicomm', 'baker', 'employed', 'social media strategist', 'saxophonist', 'chef', 'principal', 'interpreter', 'broadcaster', 'painter', 'public speaking', 'user experience', 'medicine', 'marketeer', 'production manager', 'journalist', 'filmmaker', 'microbiologist', 'referee', 'ibmer', 'associate professor', 'pharmacist', 'entertainer', 'md', 'mechanical engineer'] \\
\hline 
occupations/industry cont. & ['drummer', 'astrologer', 'physiotherapist', 'publisher', 'priest', 'director', 'future entrepreneur', 'jewelry designer', 'influencer', 'special education teacher', 'tv personality', 'barista', 'economist', 'commentator', 'editor', 'entrepreneur', 'costume designer', 'cofounder', 'kindergarten teacher', 'licensed massage therapist', 'archaeology', 'hypnotherapist', 'brand manager', 'astronomer', 'forex trader', 'private pilot', 'communications professional', 'linguist', 'emt', 'assistant', 'ordained minister', 'sound engineer', 'mechanic', 'barber', 'cmo', 'content writer', 'ecommerce', 'app developer', 'biochemist', 'insurance', 'pmp', 'math', 'enterpreneur', 'scriptwriter', 'nutritionist', 'innovator', 'publicist', 'therapist', 'clinical psychologist', 'cfa', 'it professional', 'mediator', 'future doctor', 'firefighter', 'social commentator', 'advertising', 'biotech', 'professor', 'physical therapist', 'information technology', 'rocket ship builder', 'paralegal', 'small business owner', 'stylist', 'general manager', 'copy editor', 'asst', 'supply chain', 'communications manager', 'network marketer', 'future nurse', 'consultant', 'graphic designer', 'pastor', 'social work', 'restaurateur', 'geologist', 'future lawyer', 'sculptor', 'brand ambassador', 'surgeon', 'product manager', 'recording engineer', 'welder', 'physicist', 'midwife', 'cfo', 'public speaker', 'teach', 'tourism', 'associate', 'cake decorator', 'analyst', 'dietitian', 'beautician', 'attorney', 'coaching', 'chemical engineer', 'keynote speaker', 'contentmarketing'] \\
    \hline
outdoors/nature/animals	& ['nasa', 'tree hugger', 'beekeeper', 'country boy', 'stargazer', 'sharks', '2 dogs', 'my dog', 'ecology', 'angler', 'animals', 'bunnies', 'space', 'backpacking', 'hiking', 'loves nature', 'loves animals', 'flower child', 'outdoorsman', 'mountain biker', 'nature', 'horses', 'cat', 'pandas', 'anything outdoors', 'kayaker', 'climatechange', 'animal lover', 'doglover', 'environmentalist', 'kayak', 'fishing', 'loves dogs', 'scorpion', 'environmental', 'conservationist', 'outdoors', 'animal', 'gardening', 'gardner', 'ponies', 'eagle scout', 'cats', 'sailing', 'hike', 'scuba', 'hunt', 'my cats', 'renewable energy', 'astronaut', 'pet', 'ocean lover', 'aerospace', 'horse', 'pets', 'hunting', 'pet lover', 'agriculture', 'fisher', 'birds', 'puppies', 'owls', 'dog', 'birder', 'huntin', 'surfing', 'sailor', 'hunter', 'hiker', 'farming', 'snowboarding', 'scuba diving', 'gardener', 'vet', 'ecologist', 'crazy cat lady', 'meow', 'surfer', 'veterinarian', 'mountain', 'cowboy', 'rodeo', 'shooting', 'gearhead', 'trekker', 'mountain biking', 'mountaineer', 'kitten', 'dinosaurs', '2 cats', 'naturalist', 'being outdoors', 'equestrian', 'environment', 'gator', 'skier', 'boating', 'puppy', 'turtles', 'archery', 'astrologer', 'animallover', 'fly fishing', 'climate', 'penguins', 'outdoor enthusiast', 'tiger', 'scuba diver', 'sustainability', 'astronomer', 'loves cats', 'bear', 'fisherman', 'kayaking', 'mudding', 'dog lover', 'camper', 'dogs', 'climate change', 'trail runner', 'all animals', 'wildlife', 'backpacker', 'conservation', 'scout', 'my dogs', 'surf', 'farm', 'camping']
\\
\hline
politics	& ['donald', 'president', 'make america great again', 'covfefe', 'constitutional conservative', 'political', 'trump', 'obama', 'constitution', 'vote', 'imwithher', 'warren', 'constitutionalist', 'republican', 'right', 'reagan', 'governance', 'maga', 'makeamericagreatagain', 'left', 'americafirst', 'nevertrump', 'donald trump', 'stillwithher', 'vice president', 'politics', 'vp', 'democracy', 'feelthebern', 'neverhillary', 'politician', 'realdonaldtrump', '2a', 'trumptrain', 'narendramodi', 'conservative', 'interested in politics', 'bjp', 'imran khan', 'potus', 'political junkie', 'independent', 'democrat', 'socialist', 'election2016', 'campaigner', 'public policy', 'fiscal conservative', 'uniteblue', 'libertarian', 'liberal', 'brexit', 'hillary', 'gop', 'notmypresident', 'progressive', 'trump2016', '1a', 'geopolitics', 'policy'] \\
\hline
relationship status	& ['dating', 'single', 'committed', 'poly', 'engaged', 'teamtaken', 'teamsingle', 'forever alone', 'taken', 'divorced', 'boyfriend', 'in love with', 'in a relationship', 'newlywed', 'happily taken', 'girlfriend', 'bachelor'] \\
\hline 
religion&	['believer', 'follower of jesus', 'teamgod', 'faith', 'proverbs 3:5-6', 'worship leader', 'muslimah', 'theology', 'creator', 'pjnet', 'god1st', 'jesuschrist', 'religion', 'muslim', 'my church', 'loves jesus', 'blessed', 'wiccan', 'saved', 'john 3:16', 'mohammed', 'spiritualist', 'mormon', 'child of the most high', 'jesus', 'child of the one true king', 'plans to prosper you', 'disciple', 'galatians 2:20', 'worshipper', 'loves god', 'truly blessed', 'teamblessed', 'religious', 'roman catholic', 'godisgood', 'phil 4:13', 'child of the king', 'gospel', 'godspeed', 'evangelist', 'god', 'secular', 'in god we trust', 'romans 8:28', 'catholic', 'child of god', 'visionary', 'islam', 'follow jesus', 'hindu', 'christian', 'romans 8:18', 'jewish', 'bible', 'sinner', 'witch', 'youth leader', 'church', 'spiritual', 'joshua 1:9', 'my god', 'godfirst', 'my lord', 'minister', 'allah', 'spirituality', 'romans 1:16', 'buddhist', 'chosen one', 'believe in god', 'philippians 4:13', 'declares the lord', 'preacher', 'thomasian', 'moslem', 'born again christian', 'teamjesus', 'youth pastor', 'prophet', 'follower of christ', 'godfearing', 'priest', 'born again', 'jeremiah 29:11', 'forgiven', 'christ', 'missionary', 'ordained minister', 'alhamdulillah', 'lord', 'theologian', 'pagan', 'christ follower', 'muhammad', 'amen', 'pastor', 'atheist', 'coventry', 'lucifer', 'agnostic', 'pray', 'can do all things through christ who strengthens me', 'worshiper'] \\
\hline
school/subjects	& ['class of 2018', 'prof', 'dean', 'uni student', 'ucla', 'tamu', 'lsu', "o '17", 'gamecocks', 'ubc', 'grad', 'uconn', 'buckeyes', 'first grade teacher', 'hailstate', 'college student', 'adjunct professor', 'all around nerd', 'medical student', 'fsu', 'iu', 'goblue', 'nerdy', 'h2p', 'bsn', 'faculty', 'huskers', '3rd grade teacher', 'duke', 'stem', 'geaux tigers', 'clemson', 'mizzou', 'smu', 'med student', 'lifelong learner', 'lecturer', 'international relations', 'go gators', 'sdsu', 'in college', 'university', 'nerd', 'penn', 'alumni', 'unc', 'pe teacher', 'gocougs', 'masters', 'uw', 'uniofoxford', 'engineering', 'social studies teacher', 'tcu', 'alumna', 'gatornation', 'collegestudent', 'm ed', 'educationist', 'future educator', 'science', 'msw', 'ucf', 'roll tide', 'wsu', 'uoft', 'high school', 'geeky', 'asu', 'learner', 'life-long learner', 'uga', 'mechanical engineering', '2nd grade teacher', 'buckeye', 'assistant professor', 'cornell', 'educator', 'byu', 'auburn', '8th grade', 'beardown', 'bama', 'go blue', 'school', 'buckeyenation', 'go cougs', 'boomer sooner', 'word nerd', 'penn state', 'alum', 'freshmen', 'wareagle', 'pvamu', 'nyu', 'genius', 'college', 'future teacher', 'maths', 'yale', 'rolltide', 'mit', 'utsa', 'sociology', 'univ', 'uofa', 'sophmore', 'historian', 'psu', 'perpetual student', 'scholar', 'primary school teacher', 'war eagle', 'ph d', 'geology', 'class of 2017', 'history', 'ut', 'bsc', 'studying', 'anthropology', 'kappa delta', 'graduate', 'student', 'former teacher', 'mathematician', 'educated', 'gators', 'o 2015', 'physics', 'umich', 'unlv', "class of '18", 'baylor', 'stanford', 'philosopher', 'academic', 'uf', 'junior', '5th grade teacher', 'famu', 'teacher', 'polymath', 'ncaa', 'columbia', 'fiu', 'vols', 'sophomore', 'language', 'ole miss', 'uottawa', 'polyglot', 'super junior', 'genetics', 'syracuse', 'nursing student', 'doctoral student', 'godawgs', 'preschool teacher', 'purdue', 'princeton', 'chemistry', 'pre-med', 'wolfpack', 'cal', 'teaching', '4th grade teacher', 'o 2016', 'mba', 'criminology', 'philosophy', 'hoosier', 'students', 'tarheels', 'assistant principal', 'education', 'geography', 'linguistics', 'gobucks', 'psych', 'principal', 'geek', 'class of 2016', 'goducks', 'research', 'edu', 'usc', 'associate professor', 'huge nerd', 'freshman', 'special education teacher', 'wvu', 'economics', 'alumnus', 'kindergarten teacher', 'dawgs', 'tar heel', 'go dawgs', 'biology', 'uva', 'tutor', 'badgers', 'mpa', 'hookem', 'it student', 'go vols', 'terps', 'phd', 'mathematics', 'neuroscience', 'professor', 'notre dame', 'econ', 'ksu', 'msu', 'homeschooler', 'aggie', 'varsity cheer', 'full time student', 'mad scientist', 'ncat', 'life long learner', 'sci', 'psychology', 'always learning', 'teach', 'knowledge seeker', 'ucberkeley', 'syracuseu', 'learning new things', 'senior'] \\
\hline
sex	& ['bdsm', 'sex', 'camgirl', 'playboy', 'porn', 'milf', 'tits', 'slut', 'boobs', 'horny', 'nsfw', '18+ only', 'kinky', 'sexy', 'findom', 'pussy', 'erotica', '18+ rp', 'fetish', 'dirty'] \\
\hline
sexuality&	['gaymer', 'straight', 'queer', 'gay', 'sapiosexual', 'sexuality', 'pansexual', 'lgbt', 'lgbtq', 'bisexual', 'lesbian', 'bi'] \\
\hline
sports	& ['manutd', 'redsox', 'afc', 'safc', 'cubs', 'edtech', 'skiing', 'saints', 'nffc', 'play basketball', 'badminton', 'marathon runner', 'buckeyes', 'ac milan', 'neymar', 'victoria concordia crescit', 'ball is life', 'sporty', 'braves', 'hailstate', 'athlete', 'nufc', 'mufc', 'cristiano', 'celtics', 'golf', 'capitals', 'gospursgo', 'halamadrid', 'nhl', 'nationals', 'eagles', 'ice hockey', 'red wings', 'patsnation', 'chargers', 'maverick', 'ballet', 'horse racing', 'goblue', 'mlb', 'rower', 'coyi', 'dubnation', 'football', 'patriots', 'sixers', 'h2p', 'stlcards', 'rockets', 'rams', 'huskers', 'lakersnation', 'backpacking', 'motogp', 'cowboysnation', 'go tigers', 'baller', 'giants', 'falcons', 'jets', 'hiking', '49ers', 'ski', 'nats', 'teamlebron', 'play sports', 'raptors', 'play baseball', 'racing', 'cpfc', 'volleyball', 'lfc', 'mls', 'gopackgo', 'rangers', 'cowboynation', 'lakeshow', 'ballislife', 'kobe', 'go gators', 'mystic', 'mobility', 'xc', 'nfl', 'rider', 'taekwondo', 'field hockey', 'mountain biker', 'athletics', 'sox', 'bowling', 'workout', 'fc barcelona', 'birdgang', 'hockey', 'arsenal', 'martial arts', 'packers', 'miamiheat', 'racer', 'kayaker', 'run', 'hawks', 'gym', 'cheerleader', 'lufc', 'pistons', 'wizards', 'clippers', 'red devil', 'bucs', 'kayak', 'sport', 'nyk', 'crossfitter', 'nyr', 'habs', 'colts', 'blackhawks', 'tottenham hotspur', 'triathlon', 'cfc', 'vfl', 'jays', 'mma', 'nygiants', 'muay thai', 'cyclist', 'swim', 'trickshotter', 'softball', 'basketball', 'yogini', 'angels', 'soccer', 'sportsman', 'buckeye', 'nba', 'browns', 'ronaldo', 'go pack go', 'barca', 'redskins', 'manchester united', 'lacrosse', 'teamheat', 'buckeyenation', 'go cougs', 'boomer sooner', 'titans', 'usmnt', 'tennis', 'mariners', 'sailing', 'futbol', 'golden state warriors', 'triathlete', 'gohawks', 'gooner', 'wareagle', 'hike', 'bruins', 'vince lombardi', 'fitness', 'scuba', 'warrior', 'nike', 'pilates', 'yogi', 'raidernation', 'rolltide', 'cheerleading', 'steelers', 'bullsnation', 'sfgiants', 'loves football', 'biker', 'bodybuilder', 'steelersnation', 'mets', 'cycling', 'adidas', 'swfc', 'acmilan', 'seahawks', 'golfing', 'fitfam', 'chiefs', 'spurs', 'war eagle', 'netball', 'leafs', 'ktbffh', 'ducks', 'lifting', 'gymnastics', 'huge sports fan', 'pirate', 'mcfc', 'karate', 'ball', 'hala madrid', 'gators', 'mffl', 'teamchelsea', 'bicyclist', 'thfc', 'blue jays', 'juventus', 'play hockey', 'reds', 'surfing', 'lakers', 'milanisti', 'efc', 'go bucks', 'archer', 'sailor', 'lcfc', 'canucks', 'sabres', 'pga', 'saint', 'bulls', 'avfc', 'runner', 'hiker', 'cowboys', 'manu', 'skate', 'martial artist', 'teamcowboys', 'snowboarding', 'marathoner', 'lakernation', 'warriors', 'workouts', 'skating', 'scuba diving', 'arsenalfc', 'sports', 'vols', 'caps', 'lebron james', 'hooper', 'red sox', 'yankees', 'avid golfer', 'diver', 'goalkeeper', 'surfer', 'fcb', 'bucks', 'godawgs', 'wwe', 'ynwa', 'afl', 'broncos', 'raiders', 'real madrid', 'bodybuilding', 'big sports fan', 'skateboarding', 'ravensnation', 'indians', 'rodeo', 'skateboarder', 'powerlifting', 'shooting', 'niners', 'figure skater', 'wolfpack', 'f1', 'louis cardinals', 'loves sports', 'loves the giants', 'manchesterunited', 'teamceltics', 'mountain biking', 'climber', 'baseball', 'tottenham', 'riding', 'oilers', 'cr7', 'beachbody coach', 'tarheels', 'everton', 'pacers', 'play golf', 'bowler', 'ajax', 'weights', 'snowboard', 'packer fan', 'swimming', 'devils', 'gobucks', 'boxer', 'head coach', 'whodatnation', 'dodgers', 'equestrian', 'phillies', 'cavs', 'ncfc', 'crossfit', 'exercise', 'harvard', 'weightlifting', 'skater', 'skier', 'all sports', 'evertonian', 'season ticket holder', 'goducks', 'shooter', 'referee', 'golfer', 'archery', 'paintball', 'dive', 'ufc', 'parkour', 'cardinals', 'heatnation', 'darts', 'footy', 'wrestling', 'play softball', 'penguins', 'pats', 'bikes', 'patriotsnation', 'scuba diver', 'bills', 'dawgs', 'flyers', 'tar heel', 'go dawgs', 'go bears', 'bball', 'cricket', 'dallascowboys', 'iamunited', 'messi', 'badgers', 'play soccer', 'teamlakers', 'varsity football', 'snooker', 'saintsfc', 'hoop', 'dolphins', 'wrestler', 'avid sports fan', 'hookem', 'new england patriots', 'cristian', 'kayaking', 'bengals', 'bmx', 'realmadrid', 'go vols', 'terps', 'kickboxing', 'royals', 'biking', 'bike', 'cricketer', 'judo', 'trail runner', 'mavs', 'pirates', 'orioles', 'vikings', 'bluejays', 'running', 'aggie', 'gymnast', 'panthers', 'varsity cheer', 'swimmer', 'msdhoni', 'squash', 'track', 'fcbarcelona', 'chelsea', 'carolina panthers', 'flyeaglesfly', 'whufc', 'rusher', 'knicks', 'ncat', 'fifa', 'wolves', 'water polo', 'go hawks', 'skateboard', 'astro', 'surf', 'formula 1', 'billsmafia', 'powerlifter', 'ucl', 'liverpoolfc', 'spursnation', 'mancity', 'flythew', 'footballer', 'uswnt', 'ride bikes', 'chelseafc', 'longboarding', 'astros', 'cheer', 'teamlh', 'yoga', 'gunners', 'hoops', 'rugby', 'snowboarder', 'ravens', 'steelernation'] \\
\hline
technology/ gaming/cars & ['ps3', 'android', 'game designer', 'edtech', 'gaymer', 'gta5', 'robots', 'computer', 'webdeveloper', 'motorcycle', 'data scientist', 'it engineer', 'bmw', 'it guy', 'play xbox', 'car', 'electrical', 'motogp', 'playstation', 'coder', 'civil engineer', 'video games', 'web designer', 'play video games', 'halo', 'skyrim', 'motorbikes', 'xb1', 'trains', 'airsoft', 'vr', 'avgeek', 'engineering', 'computers', 'telecom', 'gis', 'front end developer', 'gadget freak', 'minecraft', 'mvrp', 'motorcyclist', 'engineer', 'electronics', 'nascar', 'gadget geek', 'classic cars', 'clash of clans', 'digital', 'software developer', 'tech', 'game developer', 'ai', 'mechanical engineering', 'datascience', 'cybersecurity', 'videogames', 'ps4', 'motorsports', 'fintech', 'web developer', 'battlefield', 'harley rider', 'sims', 'techy', 'network engineer', 'technophile', 'gamer', 'xboxone', 'structural engineer', 'motorcycles', 'petrol head', 'world of warcraft', 'just a gamer', 'faze', 'rpg', 'xbox', 'league', 'xbox360', 'gamertag', 'we are legion', 'data science', 'electrical engineer', 'league of legends', 'board gamer', 'bigdata', 'programming', 'nlp', 'technology', 'games', 'blackjack', 'programmer', 'play minecraft', 'audio engineer', 'virtualization', 'motorcycle enthusiast', 'ux designer', 'technician', 'it specialist', 'web development', 'gaming', 'avid gamer', 'gamedev', 'software engineer', 'minecrafter', 'play cod', 'jeeps', 'website designer', 'cinchgaming', 'petrolhead', 'ffxiv', 'sony', 'board games', 'web design', 'automation', 'webdesigner', 'teamiphone', 'techie', 'casual gamer', 'play games', 'nintendo', 'callofduty', 'madden', 'website development', 'fast cars', 'trucks', 'cars', 'forzamilan', 'mod', 'jeep', 'ibmer', 'industrial', 'electronic', 'mechanical engineer', 'dota', 'planes', 'internet', 'gadget lover', 'game dev', 'mechanic', 'motocross', 'app developer', 'gadgets', 'it student', 'bmx', 'mudding', 'microsoft', 'software', 'ibm', 'auto', 'biotech', 'information technology', 'rocket ship builder', 'technologist', 'playing video games', 'cod', 'call of duty', 'coding', 'zelda', 'motorsport', 'fazeclan', 'hacker', 'it geek', 'poker', 'gta', 'machinelearning', 'car enthusiast', 'saas', 'warcraft', 'chemical engineer', 'im a gamer', 'rv'] \\
\hline
travel	& ['nomad', 'world traveller', 'avid traveler', 'will travel', 'globetrotter', 'adventurer', 'wanderlust', 'travelling', 'globe trotter', 'road trips', 'explorer', 'travel', 'traveler', 'loves to travel', 'traveller', 'wanderer', 'traveling', 'loves travelling', 'wanderluster', 'world citizen', 'world traveler'] \\
\hline 
values &	['jai hind', 'egalitarian', 'patriot', 'capitalist', 'freethinker', 'hedonist', 'truth seeker', 'peace', 'independence', 'patriotic', 'humanity', 'hakuna matata', 'peacemaker', 'radical', 'freedom', 'honesty', 'liberty'] \\
\hline
\caption{Labelled identifiers.}
    \label{tab:labels}
    \end{longtable}

\end{document}